\documentclass[conference]{IEEEtran}
\IEEEoverridecommandlockouts
\usepackage{cite}
\usepackage{amsmath,amssymb,amsfonts}
\usepackage{algorithmic}
\usepackage{graphicx}
\usepackage{textcomp}
\usepackage{multirow}
\usepackage{float}
\usepackage{microtype}
\usepackage{booktabs}

\usepackage[table,xcdraw]{xcolor}
\def\BibTeX{{\rm B\kern-.05em{\sc i\kern-.025em b}\kern-.08em
    T\kern-.1667em\lower.7ex\hbox{E}\kern-.125emX}}

\newcommand\crule[3][black]{\textcolor{#1}{\rule{#2}{#3}}}
\definecolor{myyellow}{rgb}{0.867, 0.686, 0.275}
\definecolor{myblue}{rgb}{0.180, 0.337, 0.631}
\definecolor{myred}{rgb}{1, 0, 0.1}
\definecolor{mygreen}{rgb}{0.301, 0.612, 0.440}
\definecolor{mypurple}{rgb}{0.455, 0.443, 0.678}
\definecolor{mygrey}{rgb}{0.302, 0.302, 0.302}

\definecolor{f1green}{rgb}{0.99, 0.537, 0.173}
\definecolor{f1blue}{rgb}{0.36, 0.625, 1.0}
\definecolor{f1yellow}{rgb}{0.867, 0.686, 0.275}
\definecolor{f1orange}{rgb}{0, 0.729, 0.619}

\begin{document}

\title{Quark: An Integer RISC-V Vector Processor for Sub-Byte Quantized DNN Inference\\
}

\author{\IEEEauthorblockN{MohammadHossein AskariHemmat\IEEEauthorrefmark{1},
Théo Dupuis\IEEEauthorrefmark{1},
Yoan Fournier\IEEEauthorrefmark{1}, 
Nizar El Zarif\IEEEauthorrefmark{1}, 
Matheus Cavalcante\IEEEauthorrefmark{2}, \\
Matteo Perotti\IEEEauthorrefmark{2}, 
Frank G\"urkaynak\IEEEauthorrefmark{2}, 
Luca Benini\IEEEauthorrefmark{2}, 
François Leduc-Primeau\IEEEauthorrefmark{1},
Yvon Savaria\IEEEauthorrefmark{1}, 
Jean-Pierre David\IEEEauthorrefmark{1}}
\IEEEauthorblockA{\IEEEauthorrefmark{1}Electrical Engineering Department, École Polytechnique de Montréal, Québec, Canada}
\IEEEauthorblockA{\IEEEauthorrefmark{2}Integrated Systems Laboratory, ETH Zürich, Switzerland}
}

\maketitle

\begin{abstract}
In this paper, we present Quark, an integer RISC-V vector processor specifically tailored for sub-byte DNN inference. Quark is implemented in GlobalFoundries' 22FDX FD-SOI technology. It is designed on top of Ara, an open-source 64-bit RISC-V vector processor. To accommodate sub-byte DNN inference, Quark extends Ara by adding specialized vector instructions to perform sub-byte quantized operations. We also remove the floating-point unit from Quarks' lanes and use the CVA6 RISC-V scalar core for the re-scaling operations that are required in quantized neural network inference. This makes each lane of Quark 2 times smaller and 1.9 times more power efficient compared to the ones of Ara. In this paper we show that Quark can run quantized models at sub-byte precision. Notably we show that for 1-bit and 2-bit quantized models, Quark can accelerate computation of Conv2d over various ranges of inputs and kernel sizes. 

\end{abstract}

\begin{IEEEkeywords}
RISC-V, Vector ISA, Quantization, Machine Learning, Efficiency
\end{IEEEkeywords}

\section{Introduction}
Vector processors are gaining traction for machine learning computations, thanks to their efficiency when executing tensor operations, combined with the flexibility of fully programmable instruction processing. Many Deep Neural Network (DNN) tasks can perform well with far lower precision than is typically used for operands of generic digital signal processing computations~\cite{DBLP:journals/corr/CourbariauxB16,10.1007/978-3-319-46493-0_32,Esser2020LEARNED,xu2021recu}. Quantized models operate on much smaller processing and power budgets than their full precision counterparts, making them appealing options for edge applications. Even if using sub-byte networks can help speeding up computation in neural networks, to the best of our knowledge, there are currently no other vector processors that can efficiently store, load, and operate on sub-byte operands. 

This paper makes the following contributions: first, we modified Ara \cite{Ara2020}, a RISC-V V compliant vector processor to support operations on sub-byte operands. To achieve this, we added custom instructions to accelerate low-bitwidth operations. Indeed, the majority of quantized DNN inference is performed only in the integer domain, hence, we removed the floating-point unit from Ara. This resulted in each lane of Quark being 2$\times$ smaller and 1.9$\times$ more power efficient than the ones in Ara. On the other hand, in quantized DNNs, to recover accuracy, there is a re-scaling step after each convolution or linear layer that requires floating-point operations. In Quark, higher precision operations can be performed in the CVA6~\cite{zaruba2019cost} scalar processor. 

Second, we developed a set of custom instructions, specifically designed for running sub-byte DNNs. Then, using these custom instructions, we developed a small vector DNN runtime that provides common DNN kernels operating on sub-byte data. We show that compared to \texttt{Int8} and \texttt{FP32} kernels, the quantized kernels achieve higher performance. Finally, we have implemented Quark in GlobalFoundries'
22FDX Fully-Depleted Silicon-on-Insolator (FD-SOI) technology and provide its detailed power and area usage report.

\begin{figure*}[t]
  \centering
    \includegraphics[width=\linewidth]{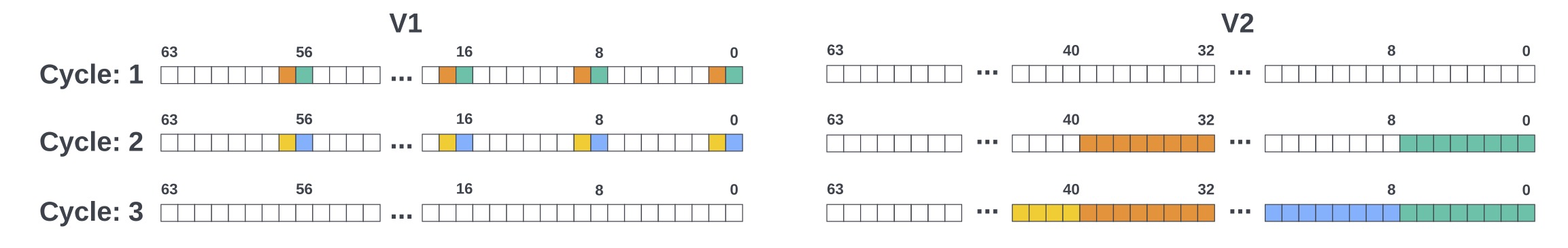}
\vspace{-.7cm}
\caption{ This figure shows the content of 8 vector elements of \texttt{v1} and \texttt{v2} when \texttt{vbitpack} instructions are executed consecutively on the same input and output vector registers. In this example, we are packing the content of \texttt{v1} into \texttt{v2} when 2 bit precision is used.  \crule[f1green]{0.2cm}{0.2cm} \crule[f1blue]{0.2cm}{0.2cm} \crule[f1yellow]{0.2cm}{0.2cm} and  \crule[f1orange]{0.2cm}{0.2cm} show different bit slices. }
\vspace{-0.2cm}
\label{fig:vbitpack_op}
\end{figure*}

The rest of the paper is organized as follows. In Section~\ref{sec:background}, we review the related works. In Section~\ref{sec:arch}, we describe the architectural changes made to Ara to make it more efficient for running quantized neural networks. In Section~\ref{sec:results}, we provide performance comparisons between Quark and Ara, and an analysis of Quark's implementation results. Finally, Section~\ref{sec:conclusion} concludes this work.

\section{Background and Related Works}
\label{sec:background}

Quantized neural networks are a well-studied research topic~\cite{gholami2021survey}. However, supporting sub-byte precision has not been as popular. In the following, we have divided the related works for these types of quantized neural networks into three different groups depending on how they are implemented: on custom accelerators, commodity CPUs and custom RISC-V Microprocessors.

\subsection{Custom Accelerators}
Since the inception of quantized neural networks, many custom accelerators based on ASIC and FPGA were proposed. Loom~\cite{sharify2018loom} supports sub-byte computation through bit-serial computation. Loom accelerates the inherent computation cost of bit-serial by using a large number of parallel bit-serial units that perform a 1-bit by 1-bit multiplication. Loom expects the input data to these parallel units to be in a bit-stream format. Because of that, Loom transposes data into a bit-stream format at each output unit, with a significant amount of overhead in area and power consumption. In~\cite{10.1145/3307650.3322255} a DNN accelerator based on Loom was proposed and implemented in 65\,nm technology, achieving 805\,GOPS when operating on 8-bit precision operands. BitFusion~\cite{sharma2018bit} and BitBlade~\cite{ryu2019bitblade} use a set of parallel 2-bit by 2-bit multipliers followed by a shift accumulation unit to support sub-byte computation. To further optimize for area and power, BitBlade~\cite{ryu2019bitblade} replaces the shift accumulation units with bitwise summation. RaPiD~\cite{9499865} is an ultra low precision accelerator for AI training and inference. RaPiD supports 16 and 8-bit floating-point and 4 and 2-bit fixed-point and was implemented in 7\,nm technology. It delivers a peak performance of 3.5\,TOPS/W in its 8-bit floating-point mode and 16.5\,TOPS/W in its 4-bit fixed-point mode. 

There are examples of FPGA-based quantized neural network accelerators as well. FINN~\cite{umuroglu2017finn} is a framework that was originally designed for the acceleration of binary neural networks. It has now evolved to a data-flow style framework with support for low precision operations for DNN acceleration. FINN can process quantized neural networks in ONNX format. Then, FINN's front-end compiler optimizes the computational graph for the FINN's back-end compiler, which uses an HLS hardware library to generate hardware supporting input quantized models. However, FINN requires the entire model to be implemented all at once on the FPGA, which limits its flexibility. 
BARVINN~\cite{barvinn_aspdac} is an FPGA based neural network accelerator that was designed to support arbitrary precision operations. It consists of a barrel RISC-V processor~\cite{9401617} used as a controller and an array of Matrix Vector Units (MVU)~\cite{8702332}. Each MVU has a set of configuration registers that can be programmed by the RISC-V processor through RISC-V's control status registers. This allows the MVU to perform a set of neural network specific operations in a pipelined fashion with any precision. The MVUs are connected to a crossbar that allows them to send the results of one MVU to another. 
Furthermore, BARVINN has a Python code generator library that can consume a quantized DNN in ONNX format and generates RISC-V code. 

\subsection{Sub-byte Precision on Commodity CPUs}
In addition to custom hardware accelerators, there have been efforts to support low-precision inference on commodity CPUs. In~\cite{cowan2020automatic}, the authors added support for bit-serial convolution in TVM to generate code for Arm devices. They exploit the vector instructions of Arm devices to overcome the computationally expensive task of formatting tensors in bit-stream format. They show that on a Cortex-A72 CPU in a Raspberry Pi 4B device, the relative speedup over 32-bit floating-point kernels for 1-bit and 2-bit precision is 6.2$\times$ and 1.7$\times$, respectively. With ULPPACK~\cite{won2022ulppack}, the authors propose a novel packing method that allows a commodity CPU to operate on sub-byte precision. In this packing method, the quantized inputs are packed into CPU's registers with enough space to allow for overflow. In~\cite{won2022ulppack}, the authors show that on a Cortex-A72 in the Raspberry Pi 4B CPU, compared to~\cite{cowan2020automatic}, ULPPACK has a better performance for 3-bit or higher precision inputs. 

\subsection{Custom RISC-V MPUs}
Dustin~\cite{9155071} is a 16-core RISC-V processor based on OpenHW Group's CV32E40P~\cite{7864441,8106976} cores. Each core has a special dot product unit that can be programmed through RISC-V's control status registers (CSRs)  to perform a dot product operation with different precision (2, 4, 8 or 16-bit operations are supported). Dustin was taped-out in 65 nm CMOS technology and achieves a peak performance of 58 GOPS and a peak efficiency of 1.15 TOPS/W. Darkside \cite{9903915} is a heterogeneous RISC-V compute cluster specifically designed for DNN inference at the edge. It has 8 RISC-V cores enhanced with 2-bit to 32-bit mixed-precision integer arithmetic. To improve performance, Darkside is equipped with a specialized datamover, a 16 bit floating point Tensor Product Engine and an accelerator for computing depthwise convolution. Darkside was implemented in 65\,nm CMOS technology and achieves a peak integer performance of 65\,GOPS on 2-bit integer kernels. 

Compared to the existing works, Quark is a vector processor that extends the RISC-V ISA by adding new instructions to accelerate packing methods for bit-serial computation.  As we will discuss in Section~\ref{sec:arch}, compared to Ara, Quark is a much slimmer processor since it does not require any vectorized floating-point operations. Compared to ASIC or FPGA based accelerators and due to the nature of general processors, Quark is very flexible and more likely to support any type of neural network that might come up in the literature in the future. Finally, none of the previous works used a vector RISC-V to provide sub-byte support for quantized DNNs.

\section{Architecture}
\label{sec:arch}

Quark's architecture is based on Ara, a modular open-source vector processor recently updated to the RISC-V V 1.0 vector ISA that is able to reach almost its theoretical peak performance when running matrix multiplication between ``large enough'' matrices~\cite{9912071}. 
The Ara system is a decoupled vector architecture composed of a CVA6, the scalar RV64GC core, and Ara, its vector engine. Each unit has independent access to the upper level of the memory hierarchy via a shared AXI bus. 

CVA6, being the only unit capable of accessing the instruction side of the memory, fetches both the scalar and vector instructions from its private L1 instruction cache, executes the former ones, and dispatches the latter ones to Ara. 
Since CVA6 is an in-order issue/commit processor that can still execute instructions out-of-order by means of a scoreboard, the dispatch of vector instructions to Ara happens non-speculatively as soon as the instruction reaches the top of the scoreboard. Then, CVA6 waits for Ara's acknowledgment and answer, before committing the instruction.
This is a fire-and-forget dispatch since Ara, if there is no result to return, acknowledges CVA6 immediately after checking that no exceptions can occur.
Ara's architecture matured through time following the RISC-V V specifications, from RVV0.5 to RVV1.0. 

Ara has been designed to accelerate a wide range of computations by means of vectorization. However, it only supports standard data types. 
In quantized neural networks, we try to avoid floating-point operations since they are computationally costly. Therefore, our design focuses on maximizing the throughput of integer computations, paying only a negligible degradation in accuracy. This allowed us to remove Ara's floating-point units and to save precious area to make room for our custom bit-serial computation instructions. In the following sub-sections, we briefly discuss these modifications.

\subsection{Bit-Serial Computation }

\begin{figure}[!h]
  \centering
    \includegraphics[width=\linewidth]{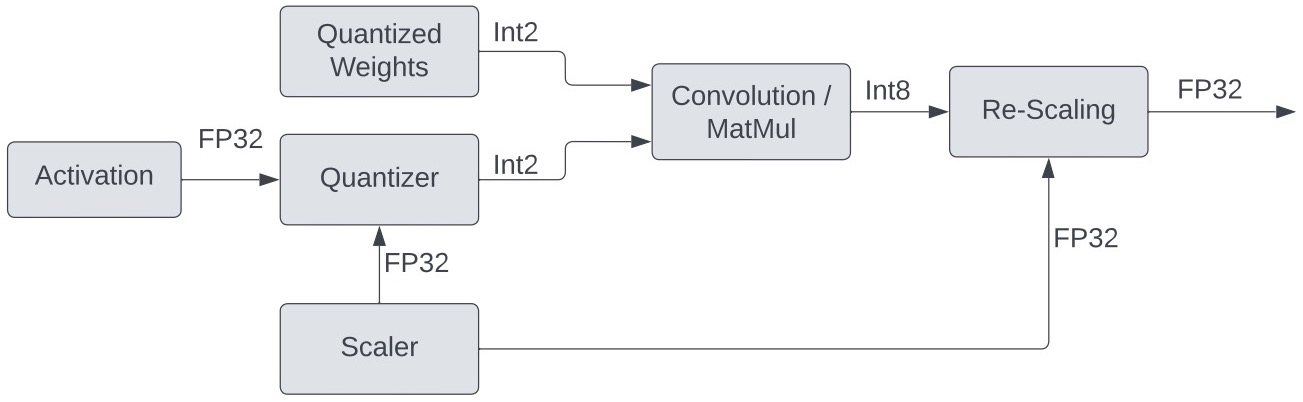}
\caption{Typical computation graph in a forward path of a quantized neural network. Here, we show the computation flow for 2 bit inference.}
\label{fig:qflow}
\end{figure}

Figure \ref{fig:qflow} illustrates a typical computation flow for quantized neural network inference. One way to accelerate the inference of neural networks is to accelerate convolutions or matrix multiplications by using low precision operations. However, as it can be seen, the rest of the computation has to happen in full precision. Since the computation complexity of convolution or matrix multiplication kernels is much higher than the rest of the operations in Figure \ref{fig:qflow}, it makes sense to accelerate them first. As discussed in Section~\ref{sec:background}, there are many ways to support low precision and particularly sub-byte operations. With Quark, this is achieved by exploiting its vector processing capacity.

As mentioned earlier, Quark extended the RISC-V vector ISA with custom instructions to speed up bit-serial computation and packing. To understand how these bit-serial computations operate, let us consider a vector-vector multiplication:

\begin{equation}
    w \cdot a = \sum_{n=0}^{N-1}\sum_{m=0}^{M-1}2^{n+m}\text{popcnt}(w_m \land a_n)
    \label{eq:bit_serial_comp}
\end{equation}

Equation~\eqref{eq:bit_serial_comp} describes a vector-vector multiplication in bit-serial format between two input vectors $w$ and $a$, where $m$ and $n$ are the precision used for $w$ and $a$, respectively. To perform this operation in a vector processor, we need logical AND, popcount, and shift-and-accumulate operators. Moreover, Equation~\eqref{eq:bit_serial_comp} expects the data to be in bit-stream format, and since this data transposition has to happen for each input of a convolution or linear layer, this data transformation should be fast to avoid making it a bottleneck to the whole computation. 
\begin{figure}[!h] 
  \centering
    \includegraphics[width=9cm]{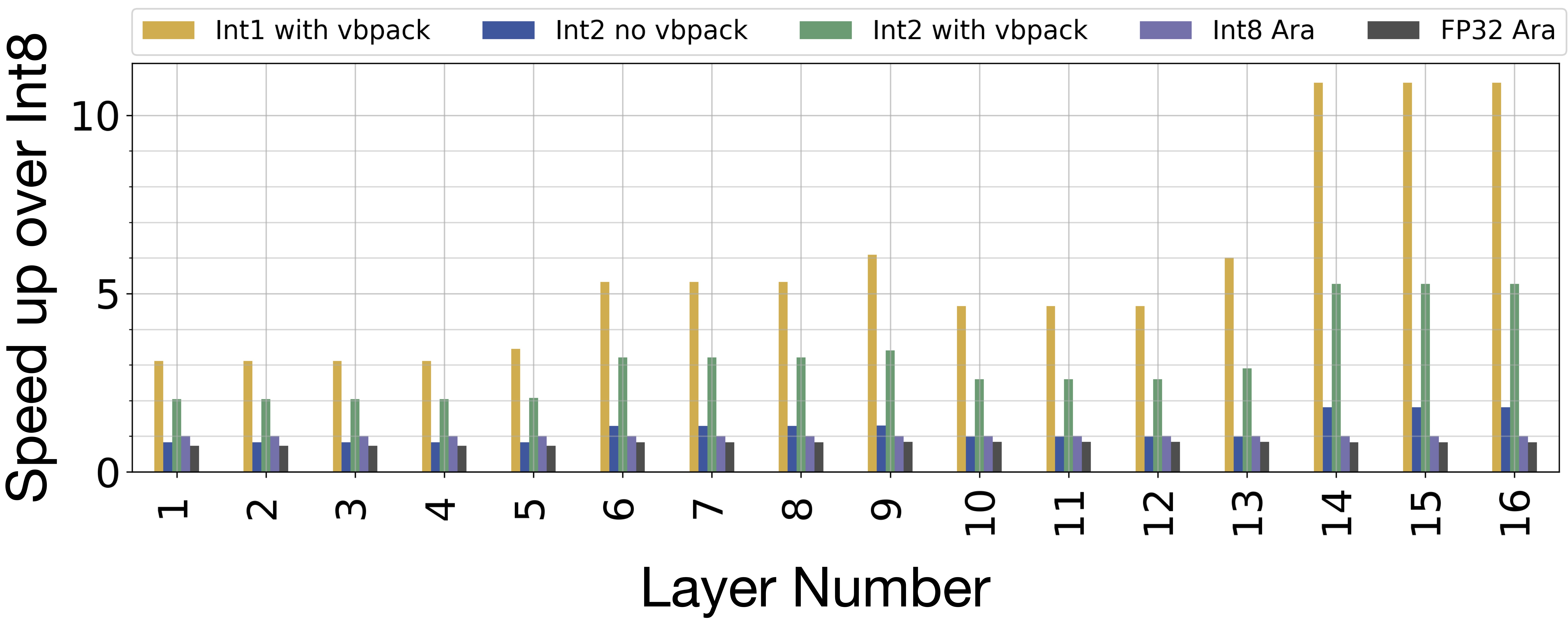}
\caption{Per layer relative speed up when running Resnet18 on CIFAR100 with batch size of 1, on Quark with \texttt{Int1} and \texttt{Int2} over Ara with \texttt{Int8}.}
\label{fig:resnet18_per_layer}
\end{figure}

\begin{figure}[!h] 
  \centering
    \includegraphics[width=9cm]{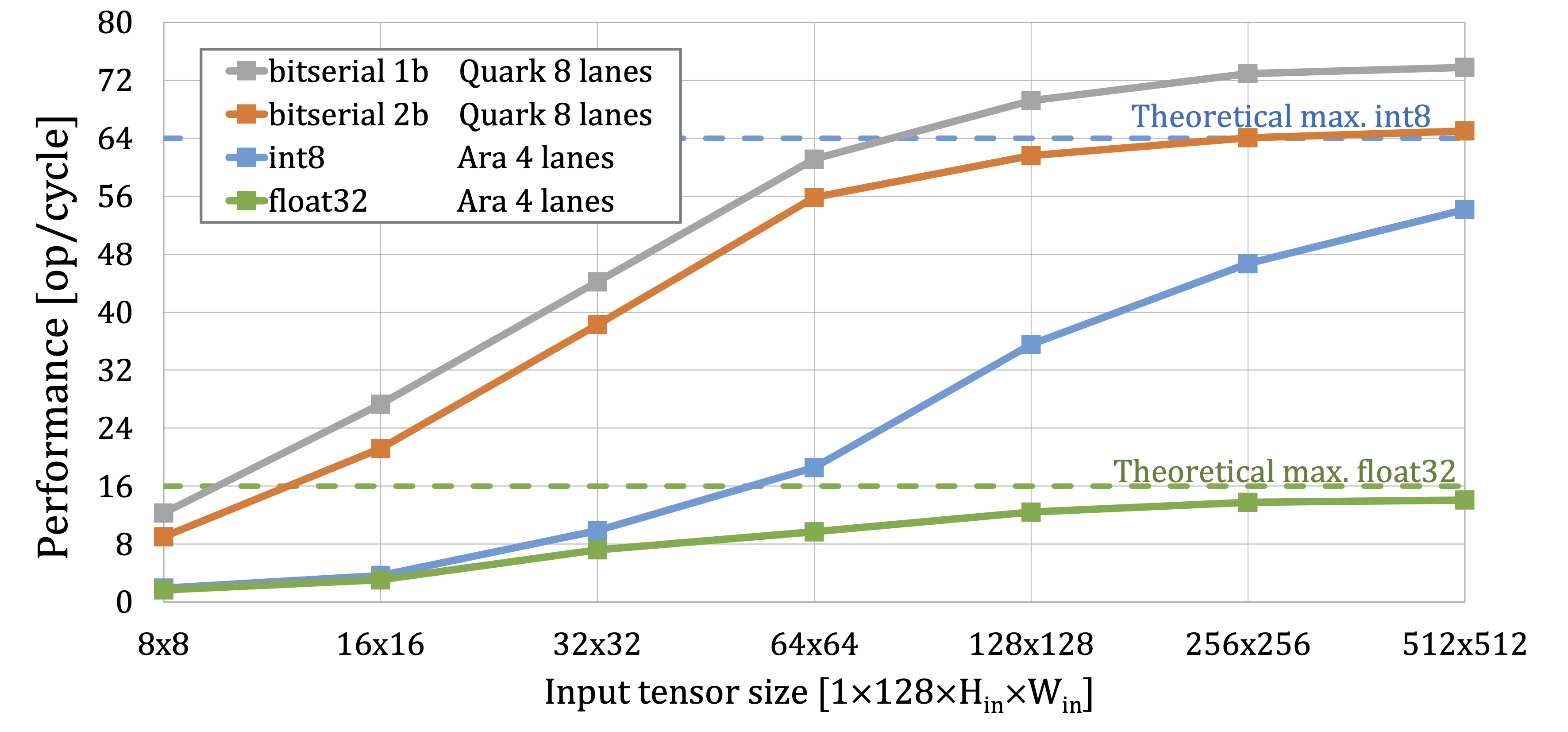}
\caption{Roofline plot for Quark with 8 lanes and Ara with 4 lanes on Conv2d with a $3\times3$ kernel.}
\label{fig:roofline}
\end{figure}

\begin{figure*}[t]
  \centering
  \begin{picture}(600,130)
\includegraphics[width=\textwidth, height=5cm]{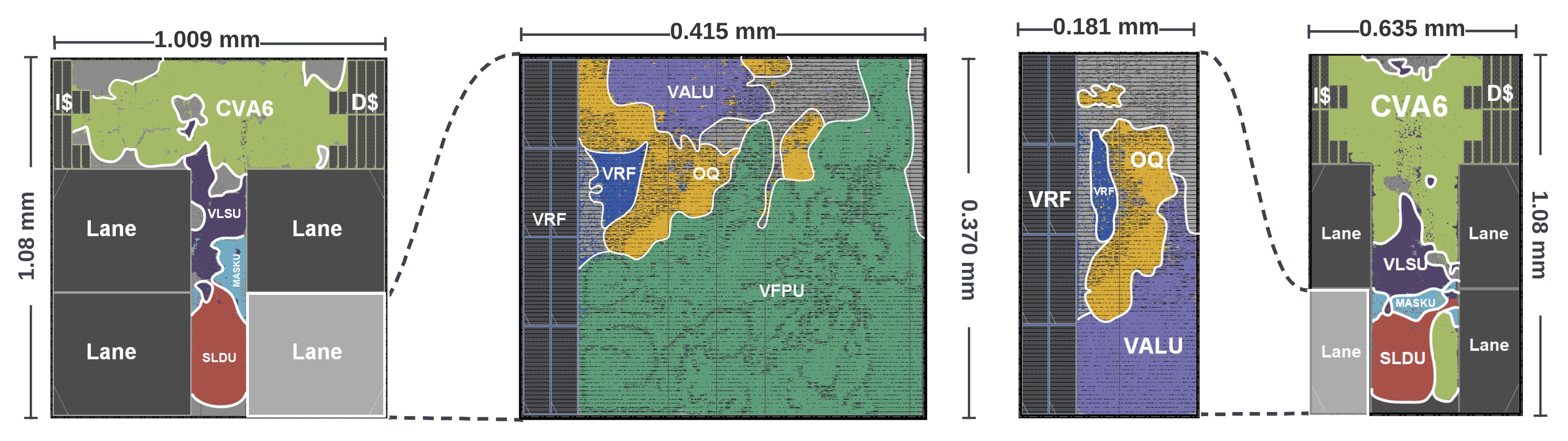}
    \put(-460,-5){Ara}
    \put(-300,-5){An Ara Lane}
    \put(-50,-5){Quark}
    \put(-170,-5){A Quark Lane}
    \end{picture}

\caption{Ara and Quark system placed-and-routed designs. \crule[myblue]{0.2cm}{0.2cm}  is vector register file related , \crule[myyellow]{0.2cm}{0.2cm}  is operand queue, \crule[mygreen]{0.2cm}{0.2cm} is the vector floating-point unit, \crule[mygrey]{0.2cm}{0.2cm} is the vector register file and \crule[mypurple]{0.2cm}{0.2cm} is the vector ALU.}
\label{fig:quark_ara_pnr}
\vspace{-0.5cm}
\end{figure*}
Although the base RISC-V vector ISA includes vectorized AND, it does not support the other operators needed. To accelerate the bit-serial computation, we added three custom instructions as follows: \texttt{vpopcnt}, \texttt{vshacc} and \texttt{vbitpack}. Our popcount operator requires an instruction able to count the number of bits at \texttt{1} for every element of the input vector; however, the base RISC-V vector ISA can only count the global number of \texttt{1}s over the entire vector length. Our special \texttt{vpopcnt} instruction performs popcount on each vector element. 
We also developed \texttt{vshacc} to fuse the shift and accumulation operations in a single instruction. Finally, we developed a custom instruction to accelerate bit-packing operations. \texttt{vbitpack} slices the elements of a vector register into bits and then packs these bits into an output register. To accumulate the bit slices of previous elements, each call to \texttt{vbitpack} shifts the target register to the left and then performs the packing operations. Figure \ref{fig:vbitpack_op} illustrates this operation in detail.

\begin{table}[htbp]
\centering

\caption{Quantization of Resnet18 Using LSQ~\cite{Esser2020LEARNED} Quantization method.}
\begin{tabular}{ccccc}
\toprule
Dataset & Model & Precision (W/A) & Accuracy & Size (MB) \\ \midrule
\multirow{4}{*}{CIFAR 100} & \multirow{4}{*}{Resnet18} & LSQ(1/1) & 57.32 & 1.45 \\ \cmidrule{3-5} 
 &  & LSQ(2/2) & 76.81 & 2.89 \\ \cmidrule{3-5} 
 &  & LSQ(8/8) & 78.45 & 10.87 \\ \cmidrule{3-5} 
 &  & FP32 & 76.82 & 42.80 \\ \bottomrule
\end{tabular}
\label{tab:lsq_quant}
\vspace{-0.35cm}
\end{table}

\section{Results and Performance Analysis}
\label{sec:results}

\subsection{Performance Analysis}
To demonstrate the benefits of adding our custom instructions, we benchmarked Ara and Quark with different kernel sizes and shapes. To do this, we developed customized bit-serial programs for conv2d, matrix multiplication, and other common kernels needed to run typical neural networks on Quark.

We benchmarked the system performance by running Resnet18 (with CIFAR100 dataset) on both Ara and Quark. To simulate, we benchmarked both Ara and Quark with Questasim 10.7c. 
For each experiment, we used CVA6's cycle CSRs to know exactly how many clock cycles each kernel takes.
To preserve the accuracy of the model, we used full precision data type for input and output layers. 
Figure~\ref{fig:resnet18_per_layer} illustrates the per-layer performance from running Resnet18 on Quark and Ara under different precisions. 
We ran Resnet18 on Ara with \texttt{Int8} and \texttt{FP32} data types and on Quark with \texttt{Int1}, \texttt{Int2} data types. To show the importance of using a specialized packing operation, we ran the \texttt{Int2} experiment 
once with RISC-V vector instructions and then with the custom \texttt{vbitpack} instruction. As Figure~\ref{fig:resnet18_per_layer} shows, running Resnet18 with \texttt{Int1} precision on Quark outperforms \texttt{Int8} on Ara in each layer. However, as  Table~\ref{tab:lsq_quant} shows, quantization of Resnet18 with 1-bit precision comes with a significant accuracy degradation. On the other hand, with 2-bit quantization, the accuracy drop is negligible, but the performance boost over \texttt{Int8} is significant. For 2-bit quantization, we measured the performance of each layer with and without 
\texttt{vbitpack} instruction. Finally, although \texttt{Int2} without the \texttt{vbitpack} instruction performs better than \texttt{Int8} on average, the improvement is not significant. On the other hand, Figure \ref{fig:resnet18_per_layer} shows that \texttt{Int2} with \texttt{vbitpack} instruction outperforms \texttt{Int8} by an average of $5.67 \times$.

\subsection{Physical Implementation Results}
We implemented Quark and Ara with GlobalFoundries' 22FDX FD-SOI technology. For both Quark and Ara, we used 4 lanes, a vector length of 4096\,bits (i.e., 16\,KiB of Vector Register File), and an equally-configured CVA6. We used Synopsys Design Compiler 21.06 for synthesis and Cadence Innovus 21.14 for physical implementation. In Table \ref{tab:impl}, we report Innovus's static power analysis for both Quark and Ara, in typical operating conditions. Figure \ref{fig:quark_ara_pnr} illustrates the placed and routed design for both Ara and Quark. Based on the results in Figure \ref{fig:quark_ara_pnr} and Table \ref{tab:impl}, each Quark lane is about $2.3 \times$ smaller and consumes $1.9 \times$ less power than Ara. Nevertheless, both cores can operate at the same 1 GHz clock frequency. Finally, Figure \ref{fig:roofline} illustrates the roofline plot for Quark and Ara on \texttt{conv2d} with a $3\times3$ kernel. In this graph, considering the same power and area budget for Ara and Quark as shown in Table \ref{tab:impl}, Quark outperforms Ara in all the input tensor sizes. 

\begin{table}[htbp]
\centering
\caption{Physical Implementation of Ara and Quark}
\begin{tabular}{rl@{\hskip 2em}ll}
\toprule
 & Ara & \multicolumn{2}{l}{Quark} \\\cmidrule(lr){2-2} \cmidrule{3-4}
Number of Lanes & 4 & 4 & 8 \\ 
VRF Size [KiB] & 16 & 16 & 32 \\ 
Lane Cell Area [mm$^2$] & 0.120 & 0.051 & 0.046  \\ 
Die Area [mm$^2$] & 1.09 & 0.69 & 1.09 \\ 
TT Frequency [GHz] & 1.05 & 1.05 & 1.00 \\
Core power per lane [mW] & 229 & 119 & 97 \\ \bottomrule

\end{tabular}
\label{tab:impl}
\end{table}

\section{Conclusion }
\label{sec:conclusion}

This paper introduced Quark, an integer RISC-V vector processor based on Ara. It extends the RISC-V vector ISA by adding custom instructions for sub-byte computations in DNNs. We modified Ara by removing the floating-point unit and adding custom instructions to accelerate bit-serial computations. Our simulation shows that on Resnet18 with 1-bit and 2-bit precision, Quark is 5.7x and 3.5x faster than Ara with 8-bit precision on average for each kerenel.
Finally, our implementation results of Quark in GlobalFoundries' 22\,nm technology show that Quark's lanes are 2.3x smaller than the corresponding ones in Ara.

\section{Acknowledgements}

The authors thank CMC Microsystems and Global Foundries for access to design tools and technologies. This research was funded by CMC Microsystems, Mitacs and the OpenHW Group. 

\bibliographystyle{IEEEtran}
\bibliography{IEEEabrv,refs}

\end{document}